\documentclass{article}

\usepackage{arxiv}

\usepackage[utf8]{inputenc} 
\usepackage[T1]{fontenc}    

\usepackage{hyperref}
\usepackage{graphicx}
\usepackage{multirow}
\usepackage{booktabs} 
\usepackage{cite}

\title{HiSA-SMFM: Historical and Sentiment Analysis based Stock Market Forecasting Model}

\date{} 					
\author{ \href{https://orcid.org/0000-0003-3746-6034}{\includegraphics[scale=0.06]{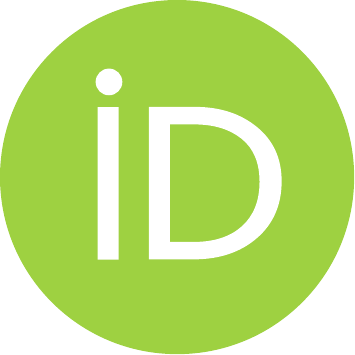}\hspace{1mm}Ishu Gupta*}
	\\
	Cloud Computing Research Center\\
	Department of Computer Science and Engineering\\ 
	National Sun Yat-sen University\\
	Kaohsiung, Taiwan\\
	\texttt{ishugupta23@gmail.com} \\
	\And
	Tarun Kumar Madan \\
	Department of Computer Applications\\
	National Institute of Technology\\
	Kurukshetra, India \\
	136119\\
	\texttt{tarunmadan9@gmail.com}
	\And
	Sukhman Singh\\
	Department of Computer Applications\\
	National Institute of Technology\\
	Kurukshetra, India \\
	136119\\
	\texttt{sukhman.07singh@gmail.com}
	\And
	\href{https://orcid.org/0000-0002-8053-5050}{\includegraphics[scale=0.06]{orcid.pdf}\hspace{1mm}Ashutosh Kumar Singh} \\
	Department of Computer Applications\\
	National Institute of Technology\\
	Kurukshetra, India \\
	136119\\
	\texttt{ashutosh@nitkkr.ac.in} \\
}

\renewcommand{\shorttitle}{\textit{HiSA-SMFM: Historical and Sentiment Analysis based Stock Market Forecasting Model }}

\hypersetup{
	pdfsubject={q-bio.NC, q-bio.QM},
	pdfauthor={David S.~Hippocampus, Elias D.~Striatum},
	pdfkeywords={First keyword, Second keyword, More},
}

\begin{document}
	\maketitle
	
	\begin{abstract}
		One of the pillars to build a country's economy is the stock market. Over the years, people are investing in stock markets to earn as much profit as possible from the amount of money that they possess. Hence, it is vital to have a prediction model which can accurately predict future stock prices. With the help of machine learning, it is not an impossible task as the various machine learning techniques if modeled properly may be able to provide the best prediction values. This would enable the investors to decide whether to buy, sell or hold the share. The aim of this paper is to predict the future of the financial stocks of a company with improved accuracy. In this paper, we have proposed the use of historical as well as sentiment data to efficiently predict stock prices by applying LSTM. It has been found by analyzing the existing research in the area of sentiment analysis that there is a strong correlation between the movement of stock prices and the publication of news articles. Therefore, in this paper, we have integrated these factors to predict the stock prices more accurately.
	\end{abstract}

	
	\section{Introduction}
	The stock market of a country is one of the pillars of its economy \cite{Sanboon,ICCNSJapan}. How the stock market performs over the years paves the way for how the economy of a country would grow or fall \cite{MLPAM,Saxena,Arora}. Since the financial markets are too uncertain, it remains unclear that the investments would bear some profits or incur some huge losses to the investors \cite{OnILIS,IJNSA,SELI}. As a part of the economic liberalization, the stock markets plays the most important role in the financial schemes of the global corporate sector \cite{Kesharwani,GUIM-SMD,Khushbu,DT-ILIS}. The biggest question for investors is what needs to be done with respect to a particular stock, i.e. whether to purchase, sell or hold the shares for a stock \cite{Jalwa,JISE,Rajat,JCOMSS}. If the investors are able to invest in the right stocks, they will bear good profits else they may lose their investments which would be a setback for them as well as their country \cite{IJAST,Ayushi,PCS}. Hence, there’s a need to devise such prediction models which may be helpful in predicting the stock prices more accurately and more efficiently \cite{Kaur2018,Jadon,Ankit}.
	
	For stock market predictions various machine learning techniques have been used to devise prediction models, few of these techniques are SVM (Support Vector Machine) \cite{Danfeng,Shen}, Linear Regression \cite{Mitesh,Sanjiban}, MLP \cite{Zhang,Turchenko}, LSTM (Long Short Term Memory) \cite{Gao}, Random Forest  \cite{Ballings,Manojlovic}, ANN (Artificial Neural Network) \cite{Rajput,Moghaddam}, etc. The various factors that affect stock prices have been used to develop these prediction models \cite{Tiwari,Animesh,Nishad,singh2020survey}. The most predominant one out of all the various factors is the prediction using historical data \cite{Godha,Holistic,IOSR}. The historical data signified the trend of the rate that had been changing over time \cite{Sharma,Sloni,MACI,Confidentiality}. With the course of time, it is observed that historical data only isn’t sufficient to get the better accuracy of the models \cite{Kamal,kaur2017comparative,BatraGarima}. In addition to historical data, sentiment data (financial news, user tweets) must also be considered as an additional factor for enhanced efficiency of the model. It is due to the reason that the sentiment data considers the current scenarios like any natural calamity, new policy change by the government, new foreign investments, etc. All these factors affect the stock market significantly but are not reflected in the historical data \cite{Vartika,Kaur2017,IDS,CC}, thus there’s a need to use the sentiment data as well. Even though the inclusion of sentiment data in prediction models isn’t easy but the efforts have been made by measuring these to integrate the same for developing more accurate prediction models.
		
	In this regard we have proposed an efficient \textbf{Hi}storical and \textbf{S}entiment \textbf{A}nalysis based \textbf{S}tock \textbf{M}arket \textbf{F}orecasting \textbf{M}odel (\textbf{HiSA-SMFM}). The model utilizes historical as well as sentiment data for more accurate stock market prediction  which majorly concerns the Indian Stock Market. The historical data has been gathered from NSE India (National Stock Exchange) and the sentiment data has been acquired using a Twitter API, Tweepy. LSTM machine learning technique is used to integrate both these aspects of stock market. The key contributions of this paper are as follows:
	\begin{itemize}
		\item An efficient model is developed that employs both historical as well as sentiment data for efficient stock market forecasting. 
		\item The tweets from Twitter through Tweepy API are extracted and then processed using Text Blob for sentiment analysis.
		\item The important features are extracted from historical and sentiment data, integrated and trained by applying LSTM machine learning algorithm.  
		\item The experiments are conducted using 'Tata Motors Stock' to validate the model and compared with the state-of-the artwork to prove the better performance of the model.
	\end{itemize}
	
	The rest of the paper is organized as follows: Section 2 discusses the key contributions and research gap in the area. Section 3 describes the proposed methodology including the technology involved and its working. The performance of the proposed methodology is validated in section 4 that entails the experimental set-up, benchmark datasets, obtained outcome, and the comparison of the model. Finally, the conclusions and the future scope of the work is presented in section 5.
	
	\section{Related Work}
	From the literature survey, it was observed that the machine learning techniques for stock market prediction are being widely used thoroughly throughout the world. Significant work has been done throughout the world in prediction analysis. Many models of prediction have been proposed to date for forecasting the stock prices and stock market trends \cite{ZhangXi,Chen,Yachika}.
	
	Yan et al. \cite{Danfeng} portrayed that the public mood and the stock market prices have some relation between them. They tried to devise a relation between the Chinese Stock Market and the Chinese local Microblogs. They worked on SVM and Probabilistic Neural Network to make predictions. The experiments results presented that Support Vector Machine provided better accuracy than the Probabilistic Neural Network. It was observed that after using the public mood as a feature, the accuracy increased by 20\% compared to using historical data only.
	Batra et al. \cite{Batra} proposed a different way of using the SVM algorithm. They extracted the sentiment from Twitter using StockTwits via an API of python. The data extracted using the former was pre-processed for sentiment analysis and Natural Language Processing (NLP). SVM was used to predict the sentiment of each data and to classify the tweets into positive and negative tweets respectively. The obtained output was combined with the available historical data and used for the stock price prediction. It was observed that the accuracy was 76.65\% in this model.
	
	Shi et al. \cite{Zhang} used Multi-Layer Perceptron to build a prediction model for the stock market in China. They found that the sentiments play a more significant role in prediction than the historical data. They gathered their dataset using a Chinese social networking site - Xueqiu. Where, they extracted the news feeds and user posts related to stocks. Their research work was based on 3 features namely sentiment feature, stock-specific features (SSF), and stock relatedness feature. The MLP technique was compared to SVM and it was observed that MLP using stock-specific features and sentiment data proved to be better than the same being used in association with SVM.
	
	Pai et al.  \cite{Turchenko} in their research work divided their project into two phases, the first one was related to predicting the stock prices on a daily basis, and the other one was based on predicting the stock prices on a monthly basis. Data for the experiments was extracted from Yahoo Finance, through which the closing and opening prices of stocks were considered. The built model considered both sentiment data and historical data. It was observed that they achieved 70\% accuracy. It was also observed that on the considered dataset, Decision Boosted Tree gave more accurate results as compared to SVM and Linear Regression. 
	
	Gao et al. \cite{Gao} proposed LSTM. In their model, there were three layers in the neural network including the input layer, the output layer, and the hidden layer, and every unit in a layer was connected with all the units in adjacent layers. It was observed that the MAPE of LSTM was the least equal to 0.7240.
	Nelson et al. \cite{Nelson} also used LSTM to build their prediction model. They used historical data to predict the stock market price movement. Initially in their model, the acquired data was preprocessed and then later used for experiments. It was found that the use of LSTM provided an average accuracy of around 55.9\%.
	Sanboon et al. \cite{Sanboon} compared the results of LSTM with support vector machine, multilayer perceptron, decision tree, random forest, logistic regression, and k-nearest neighbors. They used high, low, opening, and closing prices as features of the input. According to the results of the algorithm, the buy and sell recommendations were given. LSTM gave more accurate results compared to the other algorithms.
	
	Although, a number of models have been proposed, few of these techniques used  historical data while other focused on sentiment analysis for prediction, still an approach can be developed to enhance the model performance since these dataset alone are not sufficient. The major drawback of the existing work is that the models used a predefined values of sentiments and which is not feasible and efficient in reality. In order to overcome this drawback, a model should be developed in which this factor can be updated dynamically as per feasibility. Further, in the models, a single feature of the historical data is considered. A model is need to be developed that can accommodate multi-features.
	In this regard, we propose an approach named HiSA-SMFM for enhanced stock market forecasting based on historical and sentiment data analysis by applying LSTM machine learning algorithm. HiSA-SMFM takes into account multi-features namely open price (historical), positive reviews (sentiment), and negative reviews (sentiment).
	
	\section{Proposed HiSA-SMFM}
   In this section, the various technologies used and working of the proposed model are described.

	\subsection{Technologies}
	\subsubsection{Long Short Term Memory (LSTM)}
	Long Short Term Memory (LSTM) \cite{Hiransha} is a machine learning technique that was developed by Hochreiter and Schmidhuber in 1997. It was introduced to overcome the flaws of its predecessor RNN (Recurrent Neural Network). The biggest drawback of RNN is that it lacks learning property when the extent in between the requirement and previous information increases, this is known as the long-term dependency problem. LSTM overcomes the long-term dependency problem as it has the ability to hold the values for both long and short duration of time.
	
	LSTM has a totally different architecture as compared to other neural network models. Considering RNN, it has a very simple feedback loop neural network design whereas LSTM consists of a memory block or a cell positioned inside a single neural network layer. Since LSTM are good at remembering information for quite a long time, it becomes the first choice for use as it enhances the accuracy of the prediction models. 
	
	In our work, we proposed a model that uses an LSTM network for prediction. Our model incorporates historical as well as sentiment data to predict future stock prices. Fig. 1 shows how the LSTM Network works, here $C_{t}-1$ is the old cell state, $C_{t}$ is present cell state, $h_{t}-1$ is an output of the previous cell, $h_{t}$ is an output of the present cell, it is input gate layer, $f_{t}$ is forget gate layer and it is output sigmoid gate layer.
	
	\begin{figure}[!ht]
		\centering
		\includegraphics[width=0.8\textwidth]{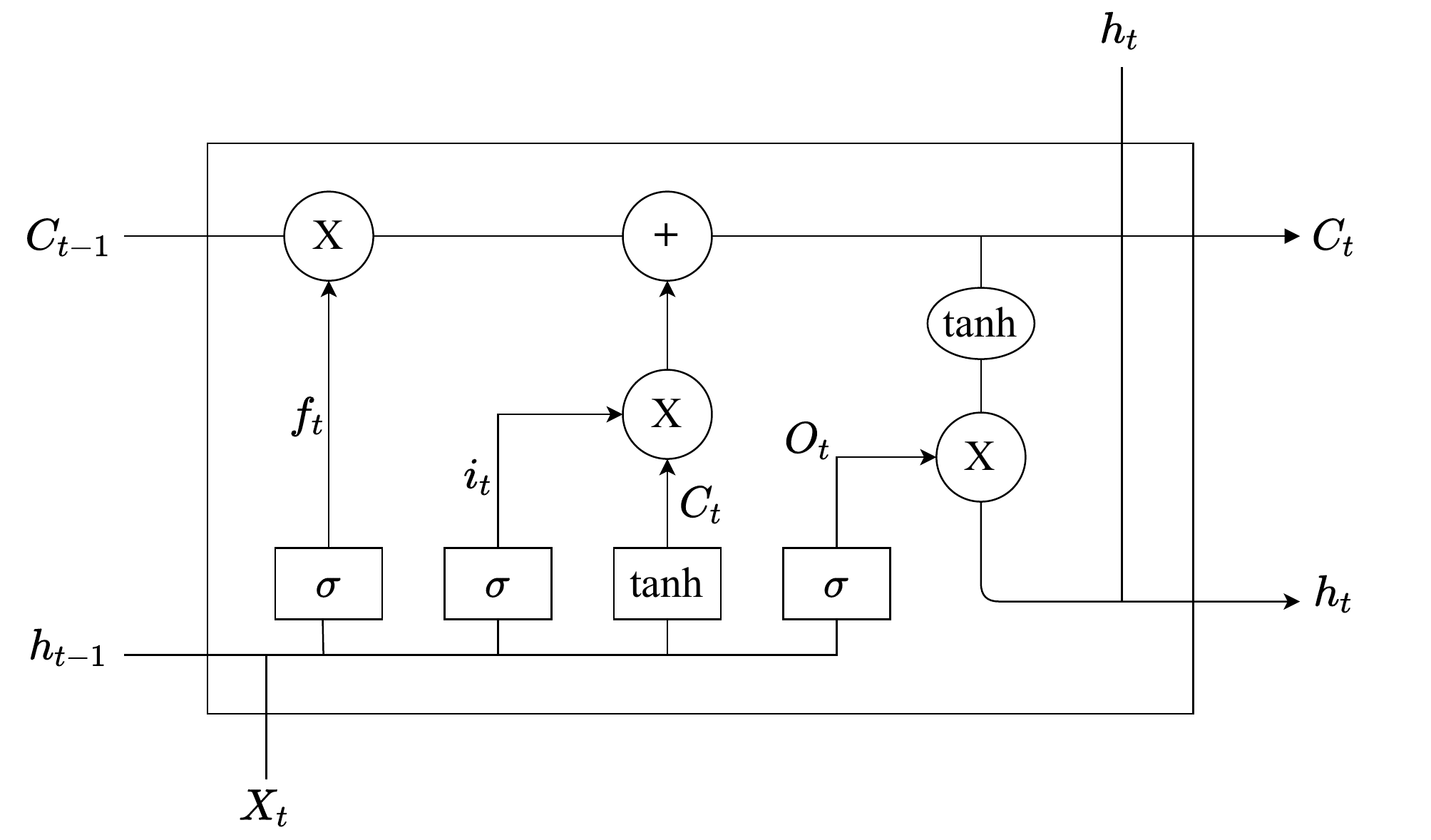}
		\caption{LSTM cell}
	\end{figure}
	
	\subsubsection{Twitter API – Tweepy}
	Twitter, one of the most popular microblogging websites plays a vital role in sentiment analysis for various fields like election results prediction \cite{Almatrafi}, cryptocurrency price prediction \cite{Abraham}, etc. Twitter provides the feature to mine their tweets data for research purposes using a Twitter API or Tweepy. The user needs to register first as a developer to gain access, then a set of keys are generated including consumer key, consumer secret, access key, and access secret. These keys are essential to link the code with the Twitter server and fetch the required data as legitimate users.
	
	\subsubsection{TextBlob}
	TextBlob is used to combine the two different kinds of judgments together, distinguish between the positive news and the negative news, and then further quantify the result of the sentiment analysis \cite{ShenAo}. It is an open-source text processing library written in Python, which provides an API for common NLP tasks such as part-of-speech tagging, noun phrase extraction, sentiment analysis, etc. Each word in the TextBlob lexicon contains the value of polarity, subjectivity and intensity. The value of polarity varies from -1  to 1. Fig. 2 portrays the working of sentiment analysis where classification of text is done in three classes positive, neutral, and negative.
	
	\begin{figure}[!ht]
		\centering
		\includegraphics[width=0.6\textwidth]{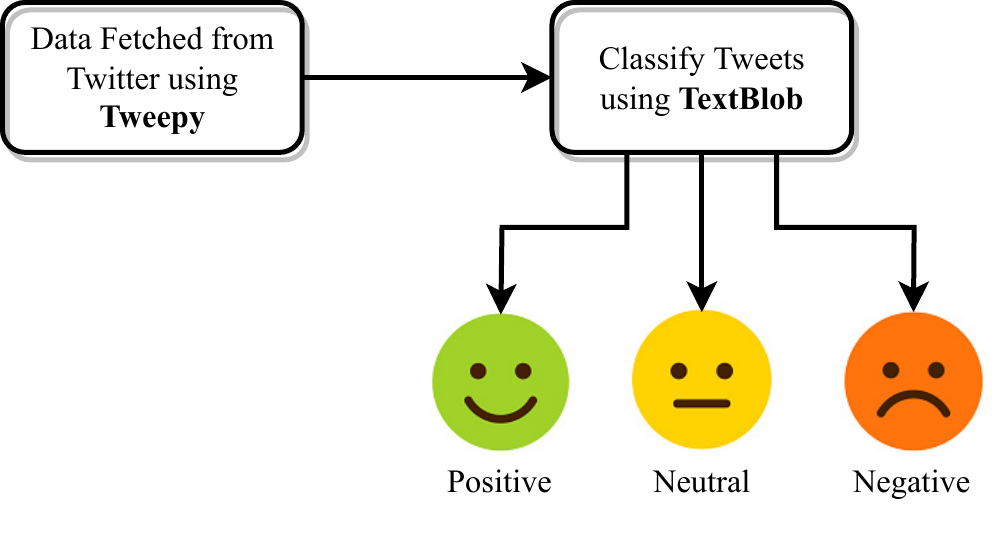}
		\caption{Working of sentiment analysis}
	\end{figure}
	
	\subsection{Working}
	The proposed model extracts the features for training the prediction model by analyzing both historical as well as sentiment data. The tweets from Twitter are extracted through Tweepy API and then processed for sentiment analysis using Text Blob. After this, we extract the historical data from NSE India. Then a model is trained for stock market prediction using stock price data and sentiment score to predict the change in the stock market. Further, the stock price prediction is performed using LSTM by adding new features into the historical dataset. The features added in the dataset are the sentiment data classified into positive and negative classes and their respective percentages. 
	
	The proposed methodology for predicting the stock market movement through sentiment analysis is carried out in 5 phases as shown in Fig. 3 that describes the working of the proposed model. The 1st phase gathers the data for historical and sentiment analysis from NSE India and Twitter respectively. In the 2nd phase, the historical data is pre-processed. The data obtained from NSE and Twitter can’t be used in the prediction model as it is, hence few amendments are essential to incorporate the data. The historical data needs to be normalized into a certain range which would maintain the trend of the market as there may be instances where the data outlies a certain boundary and might totally change the outcome. Also, the data may contain null or missing values, therefore it is important to rectify them by using the mean of the data so the trend is not hampered. Regarding the sentiment data, it is essential to use the percentage values of the obtained tweets. Once the data is classified using TextBlob, the percentage of each class is computed and added to the data. The 3rd phase classifies the data collected from Twitter using Tweepy into 3 classes, positive, negative, and neutral using TextBlob. In the 4th phase, both sentiment data and historical data are merged into one common dataset for better calculations and enhancing efficiency. The 5th phase predicts the output (stock prices) with the help of the LSTM network.
	
	\begin{figure}[!ht]
		\centering
		\includegraphics[width=0.9\textwidth]{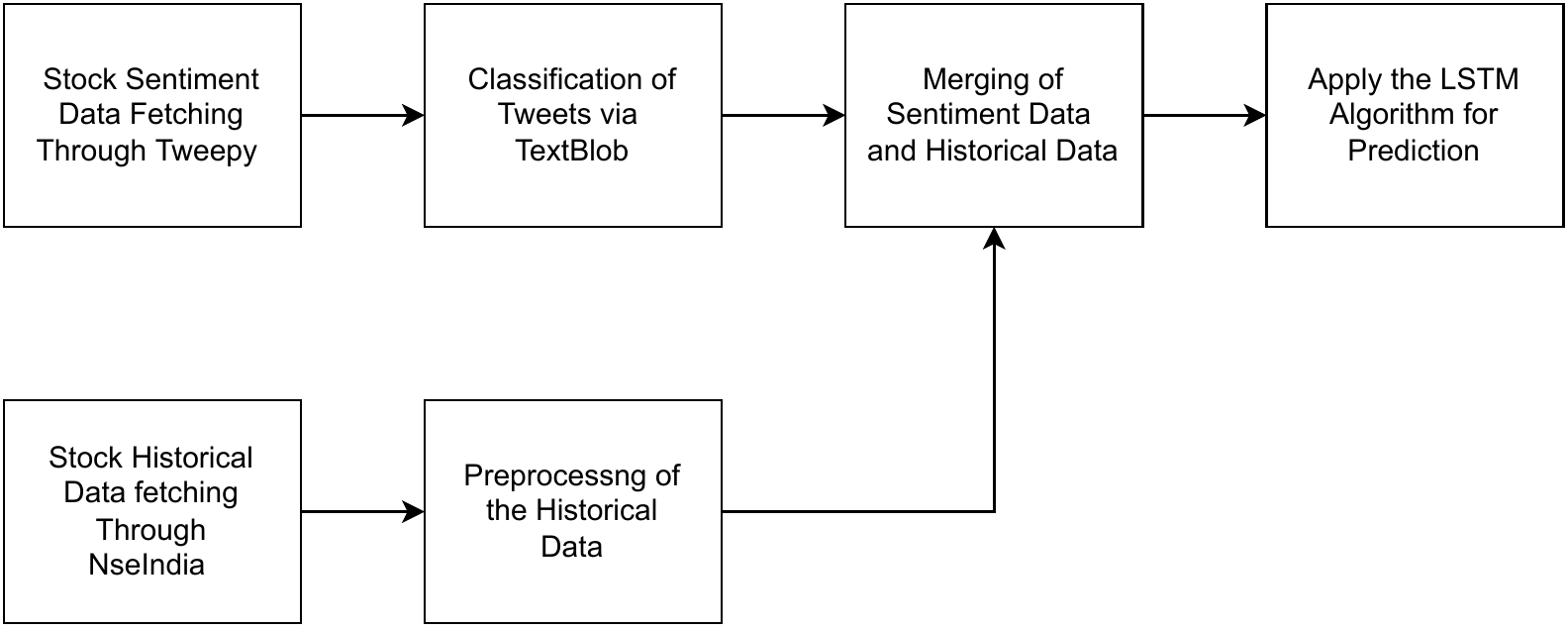}
		\caption{Working of HiSA-SMFM}
	\end{figure}
		
	\section{Performance Evaluation}
	\subsection{Experimental Set-up and Benchmark Dataset}
	The experiments for stock price prediction are conducted using 'Tata Motors dataset' to analyze the efficiency of the proposed HiSA-SMFM.
	 In HiSA-SMFM, three features namely open price (historical), positive reviews (sentiment), and negative reviews (sentiment) from the integrated sentiment and historical data are take into consideration for training the model. Furthermore, the outputs are compared with Deep Learning Prediction Model (DLPM) presented in \cite{Sanboon} for validating the performance of the model
	
	\subsubsection{Historical data}
	 The historical data from Tata Motors in the Indian Stock Market is obtained from NSE (National Stock Exchange) India website. The data set contains data for 1 year out of which data for 9 months has been used as a training dataset and the remaining data of 3 months is used as the test dataset. The dataset contains various features like date, volume, closing price, opening price, highest price, lowest price, and adjusted closing price.
	
	\subsubsection{Sentiment data}
	For the sentiment analysis part, the sentiment data has been fetched from a world-renowned microblogging website, Twitter. Using one of the Twitter API known as Tweepy, tweets can be fetched by providing certain keywords. The API searches the entire database that Twitter possesses and returns the tweets related to that keyword. Once the data has been collected, a python library known as TextBlob is used to segregate the data into three classes positive, negative and neutral. The output of Textblob is stored in the form of percentages for the various classes.
	
	\subsection{Experimental Results}
	The experiments are conducted on a different epoch sizes to measure the throughout performance of the model.
	
	\subsubsection{Real Vs Predicted Forecasting}
	Figs. 4 and 5 show the real vs predicted price with epoch size 5 of comparable DLPM \cite{Sanboon} and proposed HiSA-SMFM individually. It is found that HiSA-SMFM predicts the stock market near to the real value which is far better than prediction by \cite{Sanboon} since the training is performed by extracting multi-features from both historical as well as sentiment data in HiSA-SMFM . 
	
	\begin{figure}[!ht]
		\centering
		\includegraphics[width=0.7\textwidth]{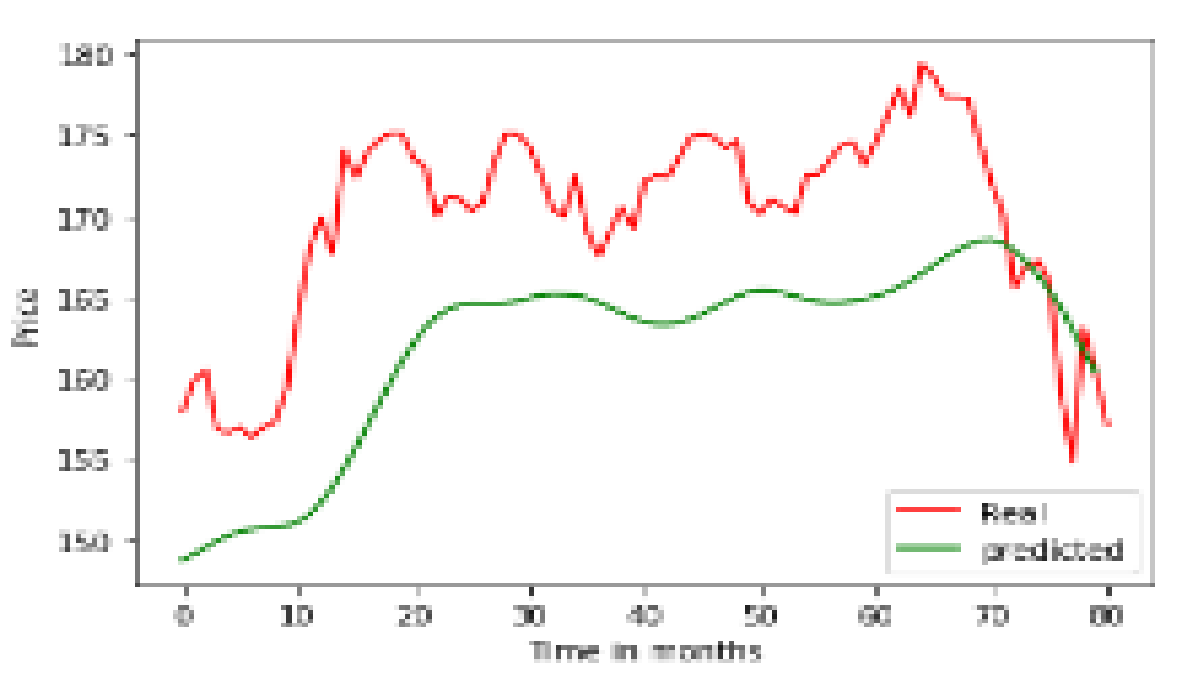}
		\caption{Real vs predicted price with epoch size 5 of DLPM \cite{Sanboon}}
	\end{figure}

	\begin{figure}[!ht]
		\centering
		\includegraphics[width=0.7\textwidth]{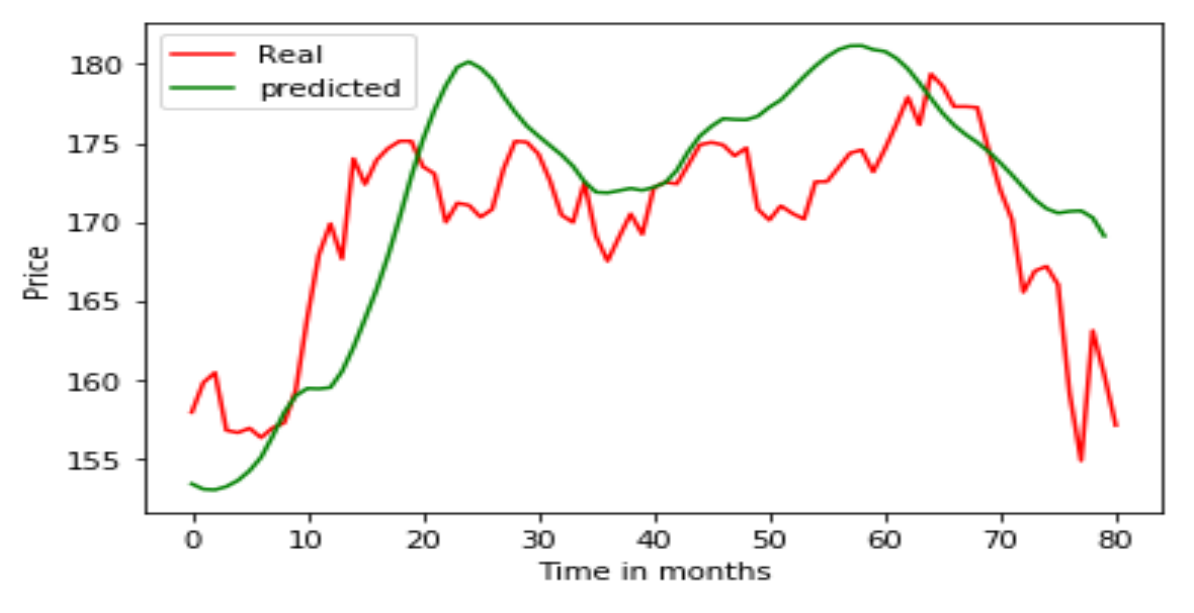}
		\caption{Real vs predicted price with epoch size 5 of HiSA-SMFM}
	\end{figure}

	Figs. 6 and 7 depict the real vs predicted price with epoch size 10 of comparable DLPM \cite{Sanboon} and proposed HiSA-SMFM respectively. It is observed that the prediction of HiSA-SMFM is nearly same as the real value which is better than DLPM \cite{Sanboon} since the integrated data of sentiment and historical is taken into account in HiSA-SMFM while historical data only is used in \cite{Sanboon}.
	
	\begin{figure}[!ht]
		\centering
		\includegraphics[width=0.7\textwidth]{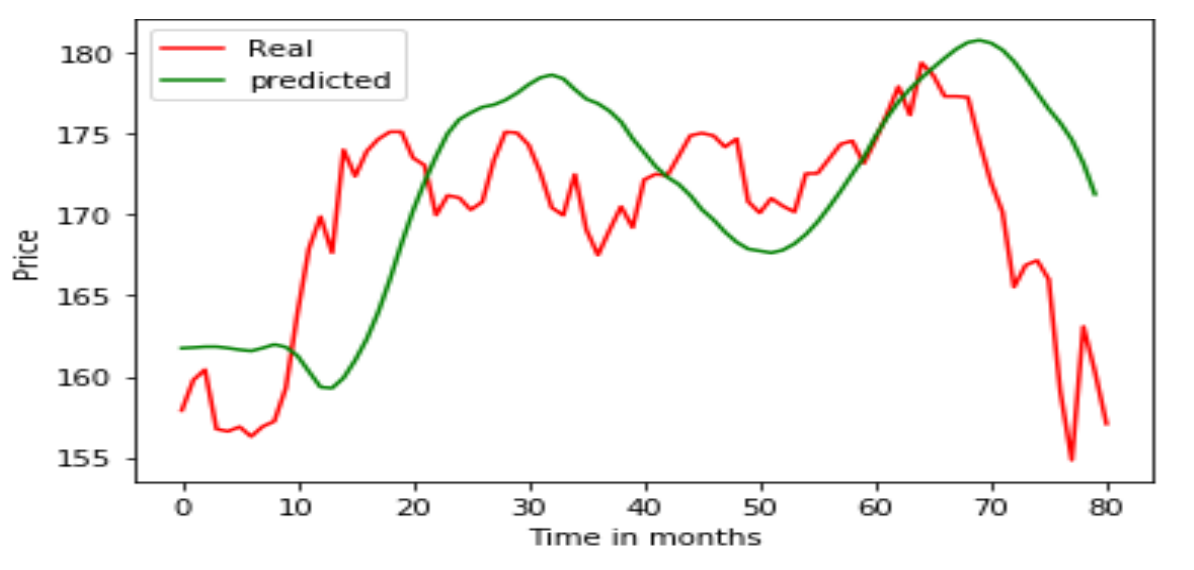}
		\caption{Real vs predicted price with epoch size 10 of DLPM \cite{Sanboon}}
	\end{figure}
	
	\begin{figure}[!ht]
		\centering
		\includegraphics[width=0.7\textwidth]{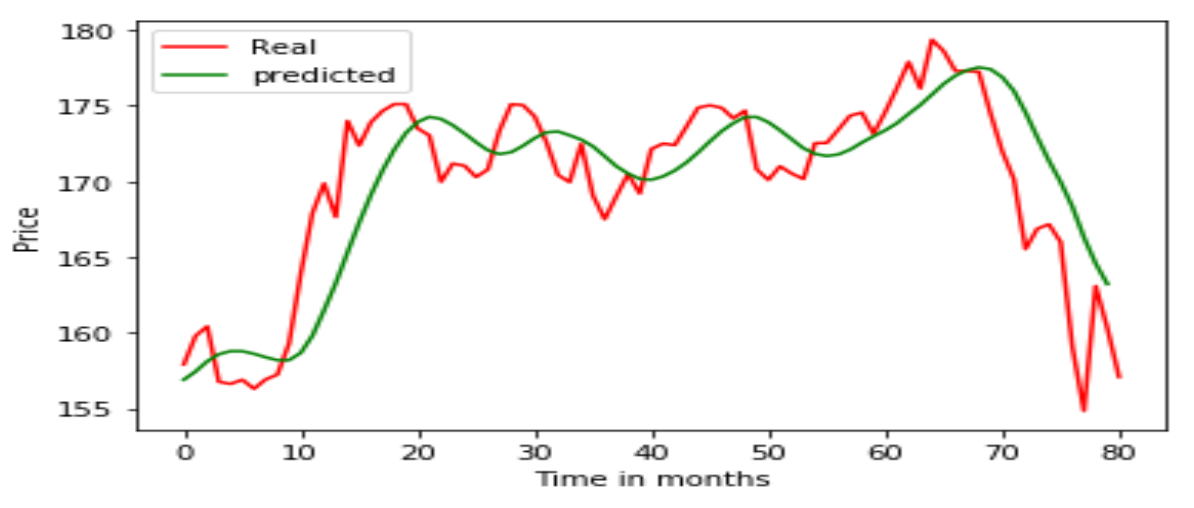}
		\caption{Real vs predicted price with epoch size 10 of HiSA-SMFM}
	\end{figure}
	
	Figs. 8 and 9 portray the real vs predicted price with epoch size 15 of comparable DLPM \cite{Sanboon} and proposed HiSA-SMFM relatively. It is observed that the prediction of HiSA-SMFM is much better than DLPM \cite{Sanboon} where the real and predicted values are far away with each other. It is due to the reason of using both historical as well as sentiment data for prediction in proposed HiSA-SMFM that extract the more accurate pattern from the input dataset.
	
	\begin{figure}[!ht]
		\centering
		\includegraphics[width=0.7\textwidth]{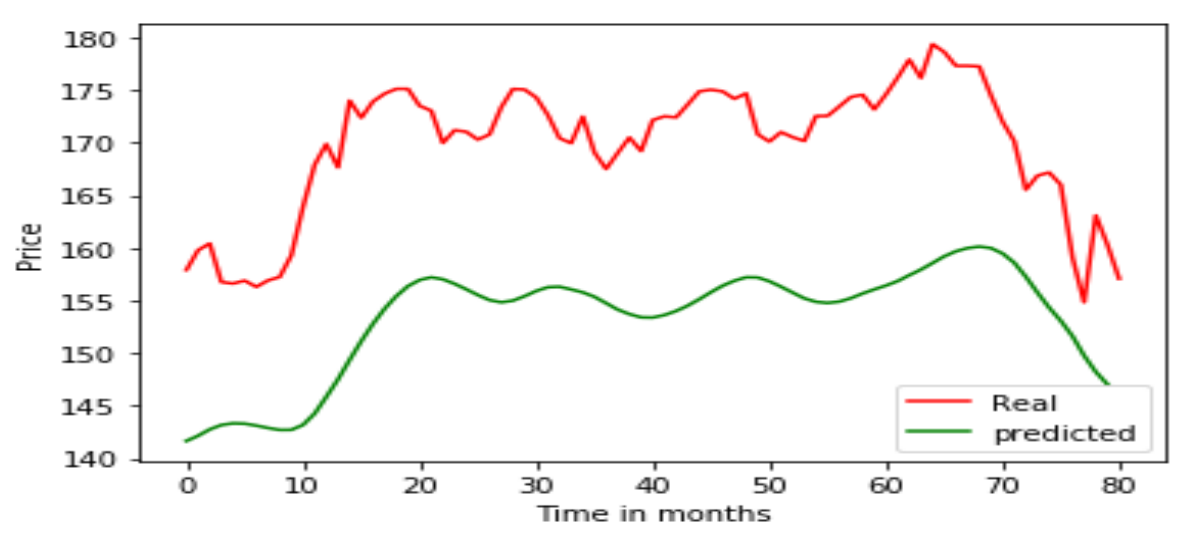}
		\caption{Real vs predicted price with epoch size 15 of DLPM \cite{Sanboon}}
	\end{figure}
	
	\begin{figure}[!ht]
		\centering
		\includegraphics[width=0.7\textwidth]{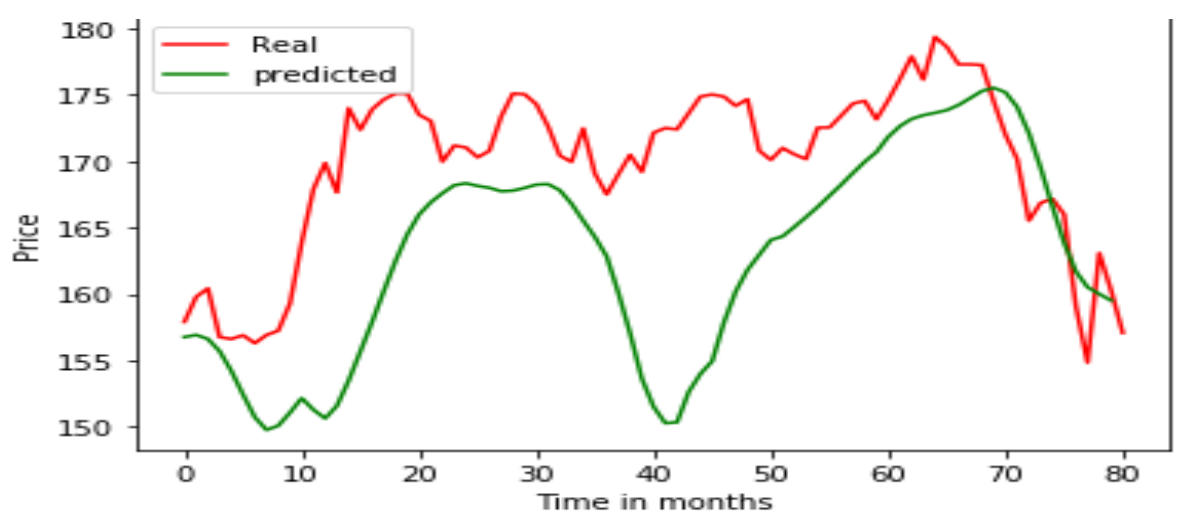}
		\caption{Real vs predicted price with epoch size 15 of HiSA-SMFM}
	\end{figure}
	
	\subsubsection{Accuracy}
	The accuracy of comparable DLPM \cite{Sanboon} and proposed HiSA-SMFM is shown in Table 1 at epoch size 5, 10, 15 followed by evaluation of average accuracy of both the models. The accuracy are 91.58\%, 95.41\% at epoch size 5, 94.56\% and 97.18\% at epoch size 10 while 83.46\% and 92.38\% at epoch size 15 respectively for DLPM \cite{Sanboon} and proposed HiSA-SMFM. The proposed model is found to be better than DLPM \cite{Sanboon}. It is found that the trend remains the same even for different epoch sizes. HiSA-SMFM ensures a significant gain of 3.82\%, 2.62\%, and 8.92\% over DLPM \cite{Sanboon} for epoch sizes 5, 10, and 15 respectively. The results show that HiSA-SMFM attains a significant improvements compared to DLPM \cite{Sanboon}. Further, DLPM \cite{Sanboon} obtains an average accuracy of 89.87\% while it is 94.99\% for HiSA-SMFM. A gain of more than 5\% is observed by the proposed model which proved HiSA-SMFM is better than state-of-the artwork DLPM \cite{Sanboon}. HiSA-SMFM ensures better performance over \cite{Sanboon} since it considers three features namely open price (historical), positive reviews (sentiment), and negative reviews (sentiment) from integrated historical and sentiment data for training the model that helps in extracting the better pattern.
	
	\begin{table} 
		\caption{Comparison of Accuracy at Various Epochs}
		\centering 
		\begin{tabular}{clc}\toprule 
			\textbf{Epoch size} &	\textbf{ Model } & \textbf{Accuracy} \\\midrule
			\multirow{2}{*}{5} &	DLPM \cite{Sanboon} & 91.59\% \\\\
			&	HiSA-SMFM & 95.41\% \\\midrule
			\multirow{2}{*}{10} &	DLPM \cite{Sanboon} & 94.56\% \\\\
			&	HiSA-SMFM & 97.18\% \\\midrule
			\multirow{2}{*}{15} &	DLPM \cite{Sanboon} & 83.46\% \\\\
			&	HiSA-SMFM & 92.38\% \\\midrule 
			\multirow{2}{*}{Average} &	DLPM \cite{Sanboon} & 89.87\% \\\\
			&	HiSA-SMFM & 94.99\% \\\bottomrule
		\end{tabular}
	\end{table}

	\section{Conclusions and Future Directions}
	As the stock market is too uncertain, the investors must invest their money after assessing the affecting factors such as public reviews, historical data, and news events from social media websites. Many researchers have tried to devise prediction models using machine learning algorithms to predict the accurate prices of stocks using various tools and techniques, but have not been able to come up with the best possible solution. In this regard, this paper proposes a novel and efficient stock market forecasting solution. The proposed model utilized the LSTM algorithm to predict the future trend of the stock market. We analyzed and combined the features of historical and sentiment data and used the LSTM algorithm to predict the future trend. We used the opening price as a prediction feature for the historical data, and Tweepy to classify the tweets acquired from Twitter. We evaluated the performance of proposed model at different epoch sizes and compared it with state-of-the artwork. It is found that the average accuracy of the proposed model is increased by more than 5\% over the existing work which is a significant improvement.
	
	 The crucial challenge faced while conducting the research work was the extraction of the sentiment data or the financial news related to certain stocks from social media websites such as Twitter followed by pre-processing of it as per the model. There are chances that people may even post fake good/bad reviews related to a particular stock so that they can either uplift or downgrade their reputation. In the future, the hybrid/fusion models can be developed to overcome the above-mentioned challenge as these models can use different techniques to acquire and process sentiment data, news events, and historical data. Also, these models have the potential to use the capabilities of their individual components that will help in forecasting more accurate results. Apart from these, a company’s balance sheet or the cash flow can be considered as an important factors to forecast the stocks. 



\begin{thebibliography}{10}
\providecommand{\url}[1]{#1}
\csname url@samestyle\endcsname
\providecommand{\newblock}{\relax}
\providecommand{\bibinfo}[2]{#2}
\providecommand{\BIBentrySTDinterwordspacing}{\spaceskip=0pt\relax}
\providecommand{\BIBentryALTinterwordstretchfactor}{4}
\providecommand{\BIBentryALTinterwordspacing}{\spaceskip=\fontdimen2\font plus
\BIBentryALTinterwordstretchfactor\fontdimen3\font minus
  \fontdimen4\font\relax}
\providecommand{\BIBforeignlanguage}[2]{{%
\expandafter\ifx\csname l@#1\endcsname\relax
\typeout{** WARNING: IEEEtran.bst: No hyphenation pattern has been}%
\typeout{** loaded for the language `#1'. Using the pattern for}%
\typeout{** the default language instead.}%
\else
\language=\csname l@#1\endcsname
\fi
#2}}
\providecommand{\BIBdecl}{\relax}
\BIBdecl

\bibitem{Sanboon}
T.~Sanboon, K.~Keatruangkamala, and S.~Jaiyen, ``{A Deep Learning Model for
  Predicting Buy and Sell Recommendations in Stock Exchange of Thailand using
  Long Short-Term Memory},'' in \emph{2019 IEEE 4th International Conference on
  Computer and Communication Systems (ICCCS)}, 2019, pp. 757--760.

\bibitem{ICCNSJapan}
I.~Gupta and A.~K. Singh, ``{A Probability based Model for Data Leakage
  Detection using Bigraph},'' in \emph{Proceedings of 7th International
  Conference on Communication and Network Security (ICCNS)}, ser. ICCNS
  2017.\hskip 1em plus 0.5em minus 0.4em\relax New York, NY, USA: Association
  for Computing Machinery (ACM), 2017, p. 1–5.

\bibitem{MLPAM}
I.~Gupta, R.~Gupta, A.~K. Singh, and R.~Buyya, ``{MLPAM: A Machine Learning and
  Probabilistic Analysis Based Model for Preserving Security and Privacy in
  Cloud Environment},'' \emph{IEEE Systems Journal}, vol.~15, no.~3, pp.
  4248--4259, 2021.

\bibitem{Saxena}
D.~Saxena, I.~Gupta, J.~Kumar, A.~K. Singh, and X.~Wen, ``{A Secure and
  Multiobjective Virtual Machine Placement Framework for Cloud Data Center},''
  \emph{IEEE Systems Journal}, pp. 1--12, 2021.

\bibitem{Arora}
U.~Arora, S.~Verma, I.~Gupta, and A.~K. Singh, ``{Implementing Privacy using
  Modified Tree and Map Technique},'' in \emph{2017 3rd International
  Conference on Advances in Computing,Communication \& Automation (ICACCA)
  (Fall)}.\hskip 1em plus 0.5em minus 0.4em\relax IEEE, 2017, pp. 1--5.

\bibitem{OnILIS}
A.~K. Singh and I.~Gupta, ``{Online Information Leaker Identification Scheme
  for Secure Data Sharing},'' \emph{Multimedia Tools and Applications},
  vol.~79, no.~41, pp. 31\,165--31\,182, November 2020.

\bibitem{IJNSA}
K.~Kaur, I.~Gupta, and A.~K. Singh, ``{A Comparative Study of the Approach
  Provided for Preventing the Data Leakage},'' \emph{International Journal of
  Network Security \& Its Applications}, vol.~9, no.~5, pp. 21--33, 2017.

\bibitem{SELI}
I.~Gupta and A.~K. Singh, ``{SELI: Statistical Evaluation based Leaker
  Identification Stochastic Scheme for Secure Data Sharing},'' \emph{IET
  Communications}, vol.~14, pp. 3607--3618, December 2020.

\bibitem{Kesharwani}
A.~Kesharwani, A.~Nag, A.~Tiwari, I.~Gupta, B.~Sharma, and A.~K. Singh,
  ``{{Real-Time Human Locator and Advance Home Security Appliances}},'' in
  \emph{Evolutionary Computing and Mobile Sustainable Networks}, vol.~53.\hskip
  1em plus 0.5em minus 0.4em\relax Singapore: Springer Singapore, 2021, pp.
  37--49, data Engineering and Communications Technologies.

\bibitem{GUIM-SMD}
I.~Gupta and A.~K. Singh, ``{GUIM-SMD: Guilty User Identification Model using
  Summation Matrix-based Distribution},'' \emph{IET Information Security},
  vol.~14, pp. 773--782, November 2020.

\bibitem{Khushbu}
Khushbu, P.~Nishad, V.~Kashyap, I.~Gupta, and A.~K. Singh, ``{{An Organized
  Study on Data Divulge Elimination and Discernment}},'' in \emph{Computer
  Networks and Inventive Communication Technologies}.\hskip 1em plus 0.5em
  minus 0.4em\relax Singapore: Springer Singapore, 2021, pp. 569--578.

\bibitem{DT-ILIS}
I.~Gupta and A.~K. Singh, ``{{Dynamic Threshold based Information Leaker
  Identification Scheme}},'' \emph{Information Processing Letters}, vol. 147,
  pp. 69 -- 73, 2019.

\bibitem{Jalwa}
S.~Jalwa, V.~Sharma, A.~R. Siddiqi, I.~Gupta, and A.~K. Singh,
  ``{{Comprehensive and Comparative Analysis of Different Files Using
  CP-ABE}},'' in \emph{Advances in Communication and Computational Technology},
  vol. 668.\hskip 1em plus 0.5em minus 0.4em\relax Singapore: Springer
  Singapore, 2021, pp. 189--198, electrical Engineering.

\bibitem{JISE}
I.~Gupta and A.~K. Singh, ``{An Integrated Approach for Data Leaker Detection
  in Cloud Environment},'' \emph{Journal of Information Science and
  Engineering}, vol.~36, pp. 993--1005, Sep. 2020.

\bibitem{Rajat}
R.~Verma, V.~Gautam, C.~P. Yadav, I.~Gupta, and A.~K. Singh, ``{A Survey on
  Data Leakage Detection and Prevention},'' in \emph{Proceedings of the
  International Conference on Innovative Computing \& Communications (ICICC)
  2020}.\hskip 1em plus 0.5em minus 0.4em\relax SSRN, Elsevier, May 2020, pp.
  1--7.

\bibitem{JCOMSS}
I.~Gupta, N.~Singh, and A.~Singh, ``{Layer-based Privacy and Security
  Architecture for Cloud Data Sharing},'' \emph{Journal of Communications
  Software and Systems (JCOMSS)}, vol.~15, no.~2, 2019.

\bibitem{IJAST}
I.~{Gupta} and A.~K. {Singh}, ``{A Framework for Malicious Agent Detection in
  Cloud Computing Environment},'' \emph{International Journal of Advanced
  Science and Technology (IJAST)}, vol. 135, pp. 49--62, Feb 2020.

\bibitem{Ayushi}
A.~Acharya, H.~Prasad, V.~Kumar, I.~Gupta, and A.~K. Singh, ``{{Host Platform
  Security and Mobile Agent Classification: A Systematic Study}},'' in
  \emph{Computer Networks and Inventive Communication Technologies},
  vol.~58.\hskip 1em plus 0.5em minus 0.4em\relax Singapore: Springer
  Singapore, 2021, pp. 1001--1010, data Engineering and Communications
  Technologies.

\bibitem{PCS}
I.~Gupta and A.~K. Singh, ``{{A Probabilistic Approach for Guilty Agent
  Detection using Bigraph after Distribution of Sample Data}},'' \emph{Procedia
  Computer Science}, vol. 125, pp. 662 -- 668, 2018.

\bibitem{Kaur2018}
K.~Kaur, I.~Gupta, and A.~K. Singh, ``{Data Leakage Prevention: E-Mail
  Protection via Gateway},'' \emph{Journal of Physics: Conference Series}, vol.
  933, p. 012013, jan 2018.

\bibitem{Jadon}
P.~Godha, S.~Jadon, A.~Patle, I.~Gupta, B.~Sharma, and A.~Kumar~Singh,
  ``{Architecture, an Efficient Routing, Applications, and Challenges in Delay
  Tolerant Network},'' in \emph{2019 International Conference on Intelligent
  Computing and Control Systems (ICCS)}.\hskip 1em plus 0.5em minus 0.4em\relax
  IEEE, 2019, pp. 824--829.

\bibitem{Ankit}
I.~Gupta, S.~Mittal, A.~Tiwari, P.~Agarwal, and A.~K. Singh, ``{TIDF-DLPM: Term
  and Inverse Document Frequency based Data Leakage Prevention Model},'' 2022.

\bibitem{Danfeng}
D.~Yan, G.~Zhou, X.~Zhao, Y.~Tian, and F.~Yang, ``{Predicting Stock Using
  Microblog Moods},'' \emph{China Communications}, vol.~13, no.~8, pp.
  244--257, 2016.

\bibitem{Shen}
\BIBentryALTinterwordspacing
S.~Shen, H.~Jiang, and T.~Zhang, ``{Stock Market Forecasting Using Machine
  Learning Algorithms},'' 2012, pp. 1--5. [Online]. Available:
  \url{http://cs229.stanford.edu/proj2012/ShenJiangZhang-StockMarketForecastingusingMachineLearningAlgorithms.pdf}
\BIBentrySTDinterwordspacing

\bibitem{Mitesh}
M.~A. Shah and C.~D. Bhavsar, ``{Predicting Stock Market using Regression
  Technique },'' \emph{Research Journal of Finance and Accounting}, vol.~6,
  no.~3, pp. 27--34, 2015.

\bibitem{Sanjiban}
S.~S. Roy, D.~Mittal, A.~Basu, and A.~Abraham, ``{{Stock Market Forecasting
  Using LASSO Linear Regression Model}},'' in \emph{Afro-European Conference
  for Industrial Advancement}, vol. 334.\hskip 1em plus 0.5em minus 0.4em\relax
  Cham: Springer International Publishing, 2015, pp. 371--381, advances in
  Intelligent Systems and Computing.

\bibitem{Zhang}
X.~Zhang, J.~Shi, D.~Wang, and B.~Fang, ``{Exploiting Investors Social Network
  for Stock Prediction in China's Market},'' \emph{Journal of Computational
  Science}, vol.~28, 11 2017.

\bibitem{Turchenko}
V.~Turchenko, P.~Beraldi, F.~De~Simone, and L.~Grandinetti, ``{Short-term Stock
  Price Prediction Using MLP in Moving Simulation Mode},'' in \emph{Proceedings
  of the 6th IEEE International Conference on Intelligent Data Acquisition and
  Advanced Computing Systems}, vol.~2, 2011, pp. 666--671.

\bibitem{Gao}
T.~Gao, Y.~Chai, and Y.~Liu, ``{Applying Long Short Term Momory Neural Networks
  for Predicting Stock Closing Price},'' in \emph{8th IEEE International
  Conference on Software Engineering and Service Science}, 2017, pp. 575--578.

\bibitem{Ballings}
M.~Ballings, D.~{Van den Poel}, N.~Hespeels, and R.~Gryp, ``{Evaluating
  Multiple Classifiers for Stock Price Direction Prediction},'' \emph{Expert
  Systems with Applications}, vol.~42, no.~20, pp. 7046--7056, 2015.

\bibitem{Manojlovic}
T.~Manojlovic and I.~Štajduhar, ``{Predicting Stock Market Trends Using Random
  Forests: A Sample of The Zagreb Stock Exchange},'' in \emph{2015 38th
  International Convention on Information and Communication Technology,
  Electronics and Microelectronics (MIPRO)}, Opatija, 05 2015, pp. 1189--1193.

\bibitem{Rajput}
G.~G. Rajput and B.~H. Kaulwar, ``{A Comparative Study of Artificial Neural
  Networks and Support Vector Machines for Predicting Stock Prices in National
  Stock Exchange of India},'' in \emph{2019 International Conference on Data
  Science and Communication (IconDSC)}, 2019, pp. 1--7.

\bibitem{Moghaddam}
A.~H. Moghaddam, M.~H. Moghaddam, and M.~Esfandyari, ``{Stock Market Index
  Prediction Using Artificial Neural Network},'' \emph{Journal of Economics,
  Finance and Administrative Science}, vol.~21, no.~41, pp. 89--93, 2016.

\bibitem{Tiwari}
P.~Tiwari, S.~Mehta, N.~Sakhuja, I.~Gupta, and A.~K. Singh, ``{{Hybrid Method
  in Identifying the Fraud Detection in the Credit Card}},'' in
  \emph{Evolutionary Computing and Mobile Sustainable Networks}, vol.~53.\hskip
  1em plus 0.5em minus 0.4em\relax Singapore: Springer Singapore, 2021, pp.
  27--35, data Engineering and Communications Technologies.

\bibitem{Animesh}
A.~Nag, A.~Kesharwani, B.~Sharma, I.~Gupta, A.~Tiwari, and A.~K. Singh,
  ``{{Potential and Extention of Internet of Things}},'' in \emph{Second
  International Conference on Computer Networks and Communication Technologies
  (ICCNCT)}, vol.~44.\hskip 1em plus 0.5em minus 0.4em\relax Cham: Springer
  International Publishing, 2020, pp. 542--551.

\bibitem{Nishad}
Khushbu, P.~Nishad, V.~Kashyap, and I.~Gupta, ``{A Classification and
  Distribution Model for Data Leakage Prevention and Detection},''
  \emph{International Research Journal of Modernization in Engineering
  Technology and Science}, vol.~3, no.~2, pp. 348--354, Feb. 2021.

\bibitem{singh2020survey}
A.~K. Singh, I.~Gupta, R.~Verma, V.~Gautam, and C.~P. Yadav, ``{A Survey on
  Data Leakage Detection and Prevention},'' in \emph{Proc. Int. Conf. Innov.
  Comput. Commun.}, 2020.

\bibitem{Godha}
P.~Godha, S.~Jadon, A.~Patle, I.~Gupta, B.~Sharma, and A.~K. Singh, ``{Flooding
  and Forwarding Based on Efficient Routing Protocol},'' in \emph{International
  Conference on Innovative Computing and Communications}, vol. 1166.\hskip 1em
  plus 0.5em minus 0.4em\relax Singapore: Springer Singapore, 2021, pp.
  215--223, advances in Intelligent Systems and Computing.

\bibitem{Holistic}
I.~Gupta and A.~K. Singh, ``{A Holistic View on Data Protection for Sharing,
  Communicating, and Computing Environments: Taxonomy and Future Directions},''
  2022.

\bibitem{IOSR}
I.~Gupta and K.~Gupta, ``{Evaluation of Intrusion Detection Schemes in Wireless
  Sensor Network},'' \emph{IOSR Journal of Computer Engineering}, vol.~18,
  no.~2, pp. 60--63, Mar-Apr. 2016.

\bibitem{Sharma}
V.~Sharma, S.~Jalwa, A.~R. Siddiqi, I.~Gupta, and A.~K. Singh, ``{{A
  Lightweight Effective Randomized Caesar Cipher Algorithm for Security of
  Data}},'' in \emph{Evolutionary Computing and Mobile Sustainable Networks},
  vol.~53.\hskip 1em plus 0.5em minus 0.4em\relax Singapore: Springer
  Singapore, 2021, pp. 411--419, data Engineering and Communications
  Technologies.

\bibitem{Sloni}
P.~Agarwal, S.~Mittal, A.~Tiwari, I.~Gupta, A.~K. Singh, and B.~Sharma,
  ``{Authenticating Cryptography over Network in Data},'' in \emph{2019
  International Conference on Intelligent Computing and Control Systems
  (ICCS)}.\hskip 1em plus 0.5em minus 0.4em\relax IEEE, 2019, pp. 632--636.

\bibitem{MACI}
A.~Acharya, H.~Prasad, V.~Kumar, I.~Gupta, and A.~K. Singh, ``{{MACI: Malicious
  API Call Identifier Model to Secure the Host Platform}},'' in
  \emph{Proceedings of the Seventh International Conference on Mathematics and
  Computing}.\hskip 1em plus 0.5em minus 0.4em\relax Singapore: Springer
  Singapore, 2022, pp. 309--320.

\bibitem{Confidentiality}
I.~Gupta and A.~K. Singh, ``{A Confidentiality Preserving Data Leaker Detection
  Model for Secure Sharing of Cloud Data using Integrated Techniques},'' in
  \emph{2019 7th International Conference on Smart Computing Communications
  (ICSCC)}.\hskip 1em plus 0.5em minus 0.4em\relax Curtin University, Sarawak
  Malaysia: IEEE, 2019, pp. 1--5.

\bibitem{Kamal}
K.~N. Kaur, Divya, I.~Gupta, and A.~K. Singh, ``{{Digital Image Watermarking
  Using (2, 2) Visual Cryptography with DWT-SVD Based Watermarking}},'' in
  \emph{Computational Intelligence in Data Mining}, vol. 711.\hskip 1em plus
  0.5em minus 0.4em\relax Singapore: Springer Singapore, 2019, pp. 77--86,
  advances in Intelligent Systems and Computing.

\bibitem{kaur2017comparative}
K.~Kaur, I.~Gupta, and A.~K. Singh, ``{A Comparative Evaluation of Data
  Leakage/Loss Prevention Systems (DLPS)},'' in \emph{Proc. 4th International
  Conference Computer Science \& Information Technology}, 2017, pp. 87--95.

\bibitem{BatraGarima}
G.~Batra, H.~Singh, I.~Gupta, and A.~K. Singh, ``{Best Fit Sharing and Power
  Aware (BFSPA) Algorithm for VM Placement in Cloud Environment},'' in
  \emph{2017 3rd International Conference on Advances in
  Computing,Communication \& Automation (ICACCA) (Fall)}.\hskip 1em plus 0.5em
  minus 0.4em\relax IEEE, 2017, pp. 1--4.

\bibitem{Vartika}
I.~Gupta, V.~Sharma, S.~Kaur, and A.~K. Singh1, ``{PCA-RF: An Efficient
  Parkinson’s Disease Prediction Model based on Random Forest
  Classification},'' 2022.

\bibitem{Kaur2017}
I.~Gupta, ``{A Comparative Study of the Approach Provided for Preventing the
  Data Leakage},'' \emph{Other Topics Engineering Research eJournal}, vol.~9,
  no.~5, September 2017.

\bibitem{IDS}
I.~Gupta and K.~Gupta, ``{Review on Intrusion Detection System Architectures in
  WSN},'' \emph{International Journal of Scientific \& Engineering Research},
  vol.~7, no.~12, pp. 111--115, Dec. 2016.

\bibitem{CC}
K.~Gupta and I.~Gupta, ``{A Comprehensive Study on Architecture, Security
  issues and Challenges in Cloud Computing},'' \emph{International Journal of
  Scientific \& Engineering Research}, vol.~7, no.~12, pp. 128--131, Dec. 2016.

\bibitem{ZhangXi}
X.~Zhang, S.~Qu, J.~Huang, B.~Fang, and P.~Yu, ``{Stock Market Prediction via
  Multi-Source Multiple Instance Learning},'' \emph{IEEE Access}, vol.~6, pp.
  50\,720--50\,728, 2018.

\bibitem{Chen}
L.~Chen, Z.~Qiao, M.~Wang, C.~Wang, R.~Du, and H.~E. Stanley, ``{Which
  Artificial Intelligence Algorithm Better Predicts the Chinese Stock
  Market?}'' \emph{IEEE Access}, vol.~6, pp. 48\,625--48\,633, 2018.

\bibitem{Yachika}
Y.~Gupta and P.~Kumar, ``{Real-Time Sentiment Analysis of Tweets: A Case Study
  of Punjab Elections},'' in \emph{IEEE International Conference on Electrical,
  Computer and Communication Technologies (ICECCT)}, 2019, pp. 1--12.

\bibitem{Batra}
R.~Batra and S.~M. Daudpota, ``{Integrating Stocktwits with Sentiment Analysis
  for Better Prediction of Stock Price Movement},'' in \emph{2018 International
  Conference on Computing, Mathematics and Engineering Technologies (iCoMET)},
  Sukkur, 2018, pp. 1--5.

\bibitem{Nelson}
\BIBentryALTinterwordspacing
A.~C. M.~P. D.~M. Q.~Nelson and R.~A. de~Oliveira, ``{Stock Market’s Price
  Movement Prediction with LSTM Neural Networks},'' in \emph{2017 International
  Joint Conference on Neural Networks (IJCNN)}, Anchorage, AK, May 2017, pp.
  1419--1426. [Online]. Available:
  \url{https://doi.org/10.1109/IJCNN.2017.7966019}
\BIBentrySTDinterwordspacing

\bibitem{Hiransha}
\BIBentryALTinterwordspacing
H.~M, G.~E.A., V.~K. Menon, and S.~K.P., ``{NSE Stock Market Prediction Using
  Deep-Learning Models},'' \emph{Procedia Computer Science}, vol. 132, pp.
  1351--1362, 2018, international Conference on Computational Intelligence and
  Data Science. [Online]. Available:
  \url{https://www.sciencedirect.com/science/article/pii/S1877050918307828}
\BIBentrySTDinterwordspacing

\bibitem{Almatrafi}
\BIBentryALTinterwordspacing
O.~Almatrafi, S.~Parack, and B.~Chavan, ``{Application of Location-Based
  Sentiment Analysis Using Twitter for Identifying Trends towards Indian
  General Elections 2014},'' in \emph{Proceedings of the 9th International
  Conference on Ubiquitous Information Management and Communication}, ser.
  IMCOM '15.\hskip 1em plus 0.5em minus 0.4em\relax New York, NY, USA:
  Association for Computing Machinery, 2015. [Online]. Available:
  \url{https://doi.org/10.1145/2701126.2701129}
\BIBentrySTDinterwordspacing

\bibitem{Abraham}
J.~Abraham, D.~Higdon, J.~Nelson, and J.~Ibarra, ``{Cryptocurrency Price
  Prediction Using Tweet Volumes and Sentiment Analysis},'' \emph{SMU Data
  Science Review}, vol.~1, no.~3, 2018.

\bibitem{ShenAo}
S.~Ao, ``{Sentiment Analysis Based on Financial Tweets and Market
  Information},'' in \emph{2018 International Conference on Audio, Language and
  Image Processing (ICALIP)}, 2018, pp. 321--326.

\end{thebibliography}
\end{document}